\documentclass[twocolumn,preprintnumbers,amsmath,amssymb,floatfix,superscriptaddress]{revtex4}

\usepackage{graphicx}
\usepackage{dcolumn}
\usepackage{float}
\usepackage{color}

\usepackage{mathrsfs} 

\begin{document}

\title{Exact correspondence between Renyi entropy flows and physical flows}

\author{Mohammad H. Ansari}
\author{Yuli V. Nazarov}
\affiliation{Kavli Institute of Nanoscience, Delft University of Technology, P.O. Box 5046, 2600 GA Delft, The Netherlands}

\date{\today}

\begin{abstract}
 We present a universal relation between the flow of a Renyi entropy and  the full counting statistics of energy  transfers.   We prove the exact relation for a flow to a system in thermal equilibrium that is weakly coupled to an arbitrary time-dependent and non-equilibrium system. The exact correspondence, given by this relation, provides a simple protocol to  quantify the flows of Shannon and Renyi entropies from the measurements of energy transfer statistics. 

\end{abstract}


\maketitle


Exact correspondences between  seemingly different concepts play important role in all fields of physics. An example is the fluctuation-dissipation theorem, which states that the linear response of a system to externally applied forces  corresponds to the system fluctuations \cite{{Nyquist},{Callan}}.   In the last decade, the  fluctuation-dissipation theorem has initiated important developments in quantum transport, quantum computation, and other similar  phenomenological theories \cite{Yuli book}. {This theorem can be extended to nonlinear responses \cite{Stratonovich} and  to full counting statistics (FCS) \cite{Tobiska}, giving more extended sets of such relations   similar to Crooks' formula \cite{relations}. }   In this paper we present a relation similar to the fluctuation-dissipation theorem that provides an exact correspondence between the flows of Renyi entropy and FCS of energy transfers.

In transport theory,   stationary flow of a physical quantity can take place from a system into an infinitely large system.  In the case a quantity is locally conserved in each system, its flow is determined only by  interaction between the two systems \cite{deutsch}.  The traditional examples include electric current, which is the flow of charge, and energy flow.  Moreover, there are  other conserved quantities which are not physical in a strict sense.   An example is the generalization of entropy by  Renyi into  $S_M=\sum_n p_n^M$, with $p_n$ being the probability to be in state $n$ and arbitrary $M>0$, \cite{renyi}.    Quantum generalization of the Renyi entropy is obviously conserved in a system under Hamiltonian evolution, with the Hamiltonian involving only the degrees of freedom of this system \cite{Nazarov}. For a system in thermal equilibrium  at temperature $T$ this entropy corresponds to the difference of free energies, i.e., $\ln S_M=F(T)-F(T/M)$. In non-equilibrium thermodynamics, the Renyi entropies have already been considered \cite{Tsallis}. They have been studied in strongly interacting systems \cite{{r1},{lieb}}, in particular spin chains \cite{{Korepin},{Abanin}}.

The Renyi entropies in quantum physics are considered unphysical,  or non-observable, due to their nonlinear dependence on  density matrix. So is the Shannon entropy, which is  derived from the Renyi entropy {$S=\lim_{M\to 1} \partial S_M/ \partial M$}, \cite{Nazarov}.  Such quantities cannot be determined from immediate measurements;   instead their quantification seems to be equivalent to determining the  density matrix. This requires reinitialization of the density matrix between many successive measurements \cite{Camprisi}.  Therefore the flows of Renyi entropy between systems $\mathcal{F}_M\equiv -d \ln S_M/dt$ are the conserved measures of non-physical quantities. The same  pertains to Shannon entropy flow \cite{Nazarov}. An interesting and non-trivial question is: Is there any relation between the flows of Renyi entropy and the physical flows?   An idea of such relation was first put forward by Levitov and Klich in \cite{levitov}, where they proposed that the Shannon entropy flow can be quantified from  the measurement of full counting statistics (FCS) of charge transfers.   The validity of this relation is restricted to vanishing temperature and obviously to the systems where interaction occurs by means of charge transfer.  This has been further elaborated in \cite{flindt}.

In this paper we present a relation which is similar in spirit. It gives a correspondence between the  flows of Renyi and Shannon entropies and the FCS of energy transfer in the limit of weak coupling.  From analysis of the previous results for entropy production in quantum point contact \cite{{levitov},{beenakker}} and more general perturbative derivations in \cite{Nazarov}, the Shannon entropy flow is known to be proportional to heat flow in the absence of external forces. This is violated in higher-order perturbation series. This implies that the exact correspondence does not hold for strong coupling limit.

 \section{Definitions and result}

  We consider two quantum systems $X$ and $Y$. We assume that the system $X$ is infinitely large and is kept in thermal equilibrium at temperature $T$.   The system $Y$ is arbitrary: it can encompass several degrees of freedom as well as infinitely many of those. It does not have to be in thermal equilibrium and in general is subject to time-dependent forces.  It is convenient to assume that these forces are periodic with period $\tau$. However this period does not enter explicitly in formulation of our result, which is also valid for aperiodic forces.  The only requirement is that there is a stationary limit of the flows of physical quantities to the system $X$. The stationary limit is defined by averaging the instant flow over the period $\tau$.  For aperiodic forces it is determined by averaging over sufficiently long time interval. 
  
The energy transfer is statistical. The \emph{FCS of energy transfers} concentrates on the probability $P(E_{tr},\mathcal{T})$  to have energy transfer of $E_{tr}$ during time interval $\mathcal{T}$,  \cite{{nazarovFCS},{kindermann}}.   In the low frequency limit of long $\mathcal{T}$ all statistical cumulants of the energy transfer are proportional to $\mathcal{T}$ and are determined from the generating function $F(\xi)=\int dE_{tr} P(E_{tr},\mathcal{T} ) \exp(i\xi E_{tr}) \approx  \exp(-\mathcal{T}   \bar{f}\left(\xi\right))$. The parameter $\xi$ is a characteristic parameter and cumulants are given by expansion of $\bar{f}(\xi)$ in $\xi$ at $\xi=0$. 

For quantification of the Renyi entropy flow we need to define an \emph{auxiliary} FCS of energy transfer.  The most {general } interaction Hamiltonian is $ \hat  H=\sum_{n}\hat{X}_{n}\hat{Y}_{n}$ with $\hat{X}_n$ being operators in the space of the system in thermal equilibrium, and  $\hat{Y}_n$ being those in the space of the arbitrary system. Let us replace $\hat{Y}_n$ with average values $\hat{Y}_n \to \langle \hat{Y}_n \rangle$.  The result in the Hamiltonian is that of the equilibrium system subject to time dependent external forces. Those induce energy transfers to the system to be characterized by a FCS.  We discuss below possible physical realization of the scheme. So we have two FCSs. We denote their generating functions   with $\bar{f_i}(\xi)$ (incoherent) and $\bar{f_c}(\xi)$ (coherent).

Our  main result  is the following exact correspondence: 
\begin{equation}\label{eq. corres}
\bar{\mathcal{F}}_{M}^{(\beta)}/M  =   \bar{f}_i^{(M\beta)}(\xi^*)-\bar{f}_c^{(M\beta)}(\xi^*) , \ \ \ \ \ \xi^*=i \beta (M-1)
\end{equation}
which indicates that the Renyi entropy flow of the order $M$ to the system kept at temperature $T=1/k_B \beta$  is exactly equal to the difference of FCS of incoherent and coherent energy transfers to the system kept at temperature $T/M$ at the fixed characteristic parameter $\xi^*$.  This relation  is valid in the limit of weak coupling, where the interaction between the systems can be treated perturbatively.

There is an obvious classical limit of the arbitrary system: all operators $\hat{Y}_n$ are just numbers  corresponding to classical forces acting on the system in thermal equilibrium. In this case the dynamics of the system is governed by the Hamiltonian in degrees of freedom of the system and therefore will be unitary.  In this case, the trace of any power of density matrix, as is used in the definition of Renyi entropy,  will not change in time: There will be no entropy flow.  This result can also be understood from the correspondence (\ref{eq. corres}): in this case $\bar{f}_i=\bar{f}_c$.

\section{Derivation of the result}

Here we discuss the proof of the exact correspondence in Eq. (\ref{eq. corres}) {in the weak coupling regime where we can restrict ourselves to the first non-vanishing order of perturbation theory.} To start with, we determine the FCS generating function of  energy transfers using a diagrammatic  representation of a pseudo-density matrix. Then we obtain the Renyi entropy flow from a  multi-contour technique, and we demonstrate the correspondence of the two.  The general formalism is illustrated by applications to two particular types of systems: the simplest quantum heat engine and a harmonic oscillator system coupled to environments.

\subsection{Full counting statistics}

Interactions between two systems influences the  statistics  of conserved quantities such as current and energy flows.  In our consideration of FCS of energy transfer,  we follow the lines of reference \cite{{nazarovFCS},{kindermann}}.  We specify it to our situation where the interaction Hamiltonian between system $X$ in thermal equilibrium and an arbitrary system $Y$ is given by $    {\hat{H}}=\sum_n \hat{X}_n \hat{Y}_n$.  The FCS of energy transfer in system $X$  during the time interval {$[0, \mathcal{T}]$} can be determined from the following  generating function: 

\begin{equation} \label{eq. gf}
{F_{\mathcal{T}} (\xi) \equiv \textup{Tr}_{_X} \tilde{\rho}_{_X}({\mathcal{T}}),}
\end{equation}
using the dynamics of the pseudo-density matrix $\tilde{\rho}$:
{ \begin{eqnarray}  
 \nonumber \label{eq. pseudo}
\widetilde{\rho}_{_X} \left(\mathcal{T} \right) & = & \rm{Tr}_{_Y}\bigg\{\bigg( \widetilde{\mathscr{T}}e^{-i\sum_{m}\int^{\mathcal{T}}_{0} dt_{2} \hat X_{m}\left(t_{2}-\frac{\xi}{2}\right)\hat Y_{m}\left(t_{2}\right)} \bigg) \bigg. \times  \\&&   \   \ \bigg.  \rho\left(0\right) \bigg( {\mathscr{T}}e^{i\sum_{n}\int_{0}^{\mathcal{T}}dt_{1} \hat X_{n}\left(t_{1}+\frac{\xi}{2}\right)\hat Y_{n}\left(t_{1}\right)}\bigg)  \bigg\}
\end{eqnarray}}
where $\widetilde{\mathscr{T}}$ ($\mathscr{T}$) denotes  (anti-) time order operator. This quantity can be rewritten as a Keldysh partition function with integral taken over a Keldysh contour. 

\begin{figure}[htbp] 
\begin{center}
\includegraphics[scale=0.28]{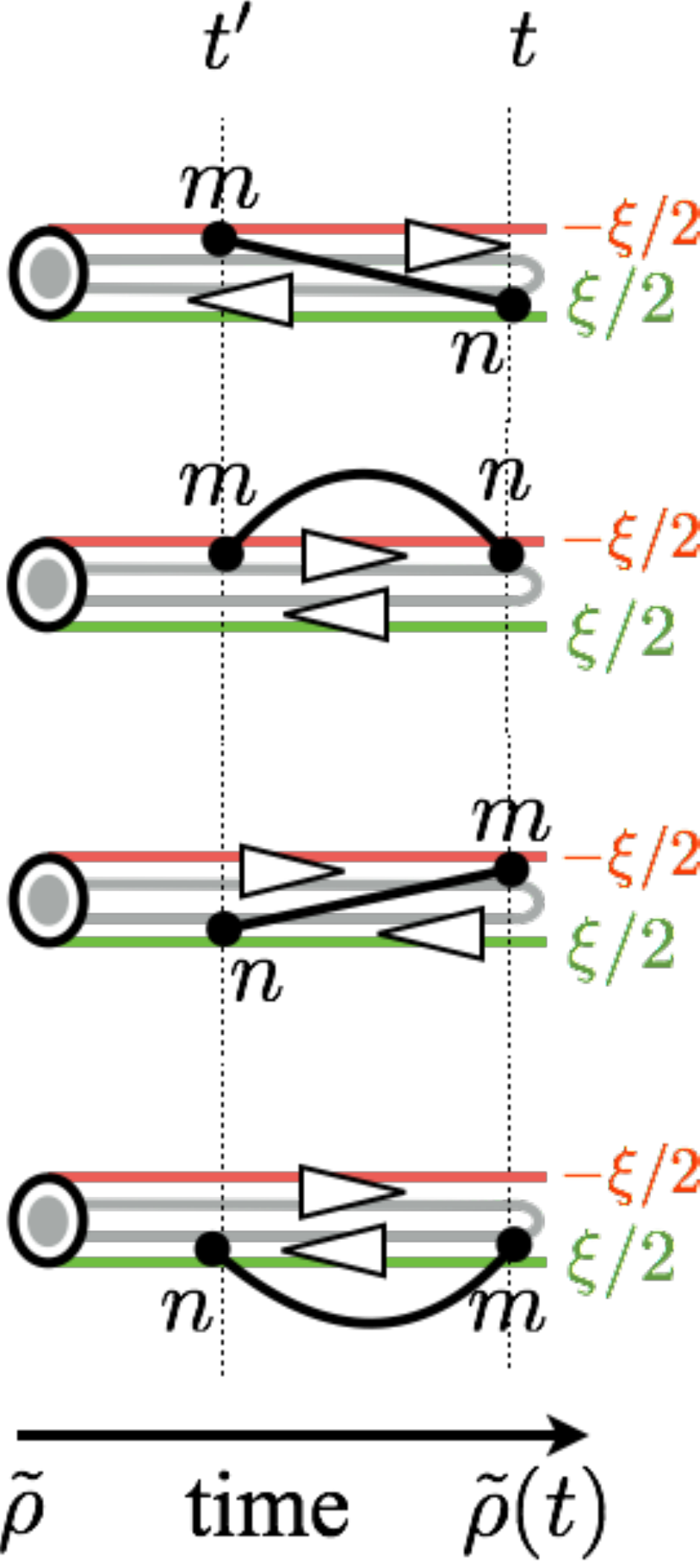}
\caption{Diagrammatic illustration of  dynamics of $\tilde{\rho}$ using  multi-contour evolution.  Arrows indicate the direction of following Keldysh contour.  In each diagram two interactions take place at vertices $m$ and $n$ and their correlator link the vertices by a solid line.  Auxiliary times  $\pm \xi/2$ shift the times when interactions act on the contour of $X$.  }
\label{fig. fcs}
\end{center}
\end{figure}

Fig. (\ref{fig. fcs}) shows that there are four possible diagrams for the evolution of  $\tilde{\rho}$ in the second order. Time moves forward from left to right.   Each diagram contains a double-contour, the outer (inner)   represents the evolution of system $X$ ($Y$).   Determining FCS for system $X$ requires to shift  $\hat X$ operators in time with $\pm \xi/2$, with $\xi$ being the \emph{characteristic parameter}. The value of the shift is opposite for forward  and backward contours corresponding to time evolution of  bra   and ket  states.

Let us consider second order perturbation for $d\tilde{\rho}_{_X}/dt$. An element of this diagram is the average $\langle X(t) X(t')\rangle $. This average is performed over the states of thermal equilibrium. We define 
\begin{equation}\label{eq. S}
S_{nm}^{(\beta)}(t-t')\equiv \langle \hat X_n(t') \hat X_m(t) \rangle. 
\end{equation} 
  The spectral density is  $S^{(\beta)}_{mn}(\omega)=\int d(t-t') \exp({i\omega (t-t')}) S_{mn}^{(\beta)} \left(t -t'\right) $. 
 Since  $t-t'$ is large in Eq. (\ref{eq. pseudo}) we can shift the lower bound $0\to -\infty$. Due to Markov approximation we can replace ${\rho}_{_X}(0)$ with ${\rho}_{_X}(\mathcal{T})$. 
 
This allows us to compute the mean value of generating function $\bar{f}(\xi)=- \rm{Tr} (d\tilde{\rho}_{_X}(t)/dt) /\rm{Tr}{\tilde{\rho}_{_X}(t)}$  averaged  over the period $\tau$. This  can be explicitly obtained from Eqs. (\ref{eq. gf}) and (\ref{eq. pseudo}) and the diagrams in Fig. (\ref{fig. fcs}):    
\begin{eqnarray} 
\label{eq. fcsdetail} 
\nonumber 
&& \bar{f}\left(\xi\right) =  - \frac{1}{\tau}\int_{0}^{\tau}dt\int_{-\infty}^{t}dt' \sum_{mn} \\&& \nonumber  
\qquad \left\{   \left[ \left\langle \hat X_{n}\left(t+\frac{\xi}{2}\right)\rho_{_X} \hat X_{m}\left(t'-\frac{\xi}{2}\right)\right\rangle \right. \right. \\ && \nonumber 
\qquad  \left.   \qquad  -\left\langle \rho_{_X} \hat X_{m}\left(t'-\frac{\xi}{2}\right) \hat X_{n}\left(t-\frac{\xi}{2}\right)\right\rangle  \right]   \\ && \nonumber 
\qquad  \qquad \qquad  \qquad \qquad \qquad \times \left\langle \hat Y_{m}\left(t'\right) \hat Y_{n}\left(t\right)\rho_{_Y}\right\rangle  \\ & & \nonumber 
\qquad  + \left[ \left\langle \hat X_{n}\left(t'+\frac{\xi}{2}\right)\rho_{_X}\hat X_{m}\left(t-\frac{\xi}{2}\right)\right\rangle \right. \\ & & \nonumber 
 \qquad  \qquad \left.  -\left\langle \hat X_{m}\left(t+\frac{\xi}{2}\right) \hat X_{n}\left(t'+\frac{\xi}{2}\right)\rho_{_X}\right\rangle \right] \\&& \nonumber 
\qquad  \qquad \qquad  \qquad \qquad \qquad  \times \bigg. \left\langle \hat Y_{m}\left(t\right) \hat Y_{n}\left(t'\right)\rho_{_Y}\right\rangle \bigg\} \\ 
\end{eqnarray}


\emph{FCS of incoherent energy transfer}: The incoherent energy transfer takes place between the system $X$ and $Y$ using the interaction Hamiltonian $\hat{H}=\sum_n \hat{X}_n \hat{Y}_n$. The correlator defined in Eq. (\ref{eq. S}) helps to simplify Eq. (\ref{eq. fcsdetail}).  After some easy steps, the generating  function of the  incoherent FCS in the probe environment becomes:
\begin{equation} 
\label{eq. fi}
\bar{f}^{(\beta)}_i\left(\xi\right)  =  - \sum_{mn} \int\frac{d\omega}{2\pi}\left(e^{-i\omega\xi}-1\right)S_{mn} ^{\left(\beta\right)}\left(\omega\right)  \mathcal{Y}_{mn}(\omega)
\end{equation}
with the definition   
\begin{eqnarray} \nonumber
\label{eq. curlyY}
&& \mathcal{Y}_{mn}\left( \omega \right)  \equiv  {\frac{1}{\tau}   \int_{0}^{\tau}dt }\int_{-\infty}^{t}dt'  \\  \nonumber & &   \bigg\{  \left\langle \hat Y_{m}\left(t'\right) \hat Y_{n}\left(t\right)  \right\rangle e^{-i\omega (t-t')}  +  \left\langle \hat Y_{m}\left(t\right) \hat Y_{n} \left( t'\right)  \right\rangle e^{i\omega (t-t') }   \bigg\} \\ 
\end{eqnarray}


\emph{FCS of coherent energy transfer}: Since a driving force is  externally applied,   another type of energy exchange is possible to take place between the driving force and the system  $X$.   In this sense, two opposite energy transfers occur between the external force and system $X$, one at  $t$ and the other at $t'$.  The two transfers are correlated from within the system $X$. One way to consider this energy transfer is to replace $\hat{Y_n}$ with the driving energy:  $\hat{Y}_n \to \langle \hat{Y}_n\rangle$. The interaction Hamiltonian is $ \hat  H=\sum_n \hat{X}_n \langle \hat{Y}_n \rangle +h.c$.   The full counting statistics of coherent energy transfers can be described in a similar way as that of incoherent energy transfer discussed above:
\begin{equation} 
\label{eq. fc}
\bar{f}_c^{(\beta)}\left(\xi\right)  =  - \sum_{mn} \int\frac{d\omega}{2\pi}\left(e^{-i\omega\xi}-1\right)S_{mn} ^{\left(\beta\right)}\left(\omega\right)  \textup{Y}_{mn}(\omega),
\end{equation}
\begin{eqnarray} \nonumber
\label{eq. textY}
  \textup{Y}_{mn}\left( \omega \right)  &\equiv& { \frac{1}{\tau}   \int_{0}^{\tau}dt }\int_{-\infty}^{t}dt'  \\ \nonumber  & &    \qquad \bigg\{  \left\langle \hat Y_{m}\left(t'\right)  \right\rangle \left\langle \hat Y_{n}\left(t\right)  \right\rangle e^{-i\omega (t-t')}\bigg. \\ &&   \qquad+\bigg.  \left\langle \hat Y_{m}\left(t\right) \right\rangle \left\langle  \hat Y_{n} \left(t'\right)  \right\rangle e^{i\omega (t-t') }   \bigg\} 
\end{eqnarray}
with  ${Y}_{mn}$ being spectral density of the forces acting on the system $X$. 

 Using the relation between $S$ and response function $\tilde{\chi}_{mn}$ (see Appendix \ref{app3} ) one can rewrite the coherent FCS in more comprehensive way:
\begin{eqnarray} 
\label{eq. fccomp}\nonumber
\bar{f}_c^{(\beta)}\left(\xi\right)  &=&  -  \int_0^\infty \frac{d\omega}{2\pi}  \sum_{mn} \tilde{\chi}_{mn}^{(\beta)} (\omega )   \textup{Y}_{mn}(\omega) \times \\ \nonumber  && \left[ \left(e^{-i\omega\xi}-1\right)  \bar{n}\left({\omega}/{T}\right)  + \left(e^{i\omega\xi}-1\right)  \left(  \bar{n}\left({\omega}/{T}\right) +1\right) \right]\\
\end{eqnarray}

Statistical  cumulants $C_n$ can be determined from the FCS generating functions from $C_n= i^n d^n\bar{f}/d\xi^n$ at $\xi=0$.

\subsection{Renyi entropy flow}

The fluctuation relations are traditionally formulated in terms of entropy production that is computed using classical states \cite{Seifert}. When it comes to quantum, the Shannonon entropy is known to be non-linear in density matrix and its change is not necessarily related to the expectation value of any operator. This problem raises a careful consideration of entropy, specially that with current technology developments entropy production in small scale systems is revealing the rich physics yet to be fully probed \cite{{experimental},{theoretical}}.

A generalization of Shannon entropy is the Renyi entropies. To evaluate the flow of Renyi entropy (R-flow) we need to use the perturbation theory for the $M$-th power of its density matrix. \cite{Nazarov} To this end, we use a multi-contour Keldysh technique. We consider $M$ copies of an isolated world. The contour for the degrees of freedom of $X$ encompasses all of the worlds and closes. This  imposes the trace over the matrix multiplication of $\rho_{_X}$. For  other degrees of freedom in $Y$, the bra and ket parts of the contours are closed within each world providing the partial trace over these degrees of freedom: that yields $\rho_{_X}=\textup{Tr}_{_Y}\rho$ for each world. The relevant diagrams are pairwise-grouped.  

The average flow of Renyi entropy during a period  {$\tau$}  is simply determined from {$\bar{\mathcal{F}}_M=(1/\tau) \int_0^{\tau} {\mathcal{F}}_M dt$}.  In the second order we expect two interactions of the form indicated above Eq. (\ref{eq. gf}).  The two interactions can be either in the same world, or in different ones.   The same-world diagrams have  been considered in \cite{Nazarov}. The different-world diagrams contain contributions from quantum coherence terms and are present when driving force is applied. We studied the contribution of the quantum coherence on the Renyi entropy flows in \cite{AN14}.

\begin{figure}[htbp]
\begin{center}
\centerline{\includegraphics[width=0.5\linewidth]{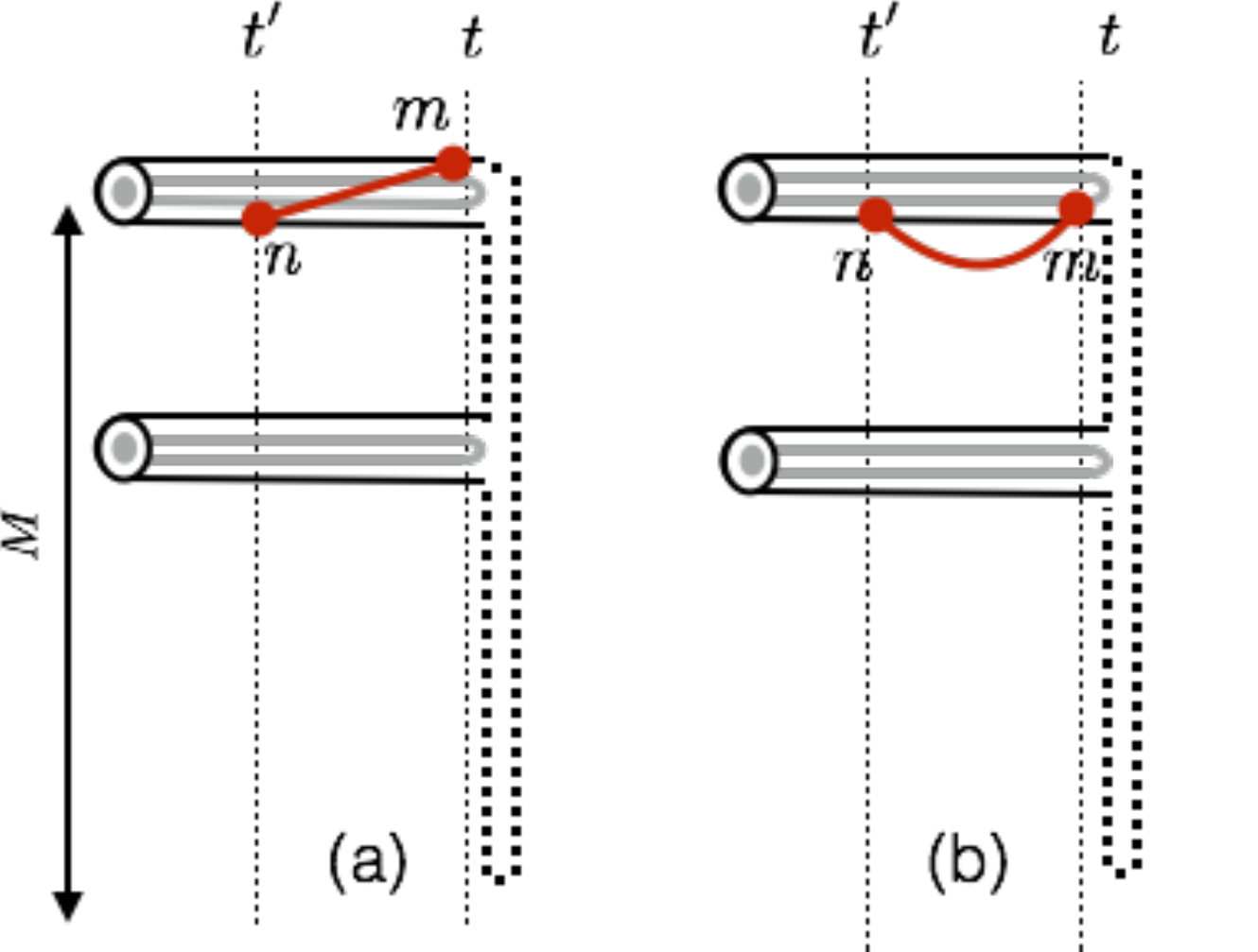}\includegraphics[width=0.5\linewidth]{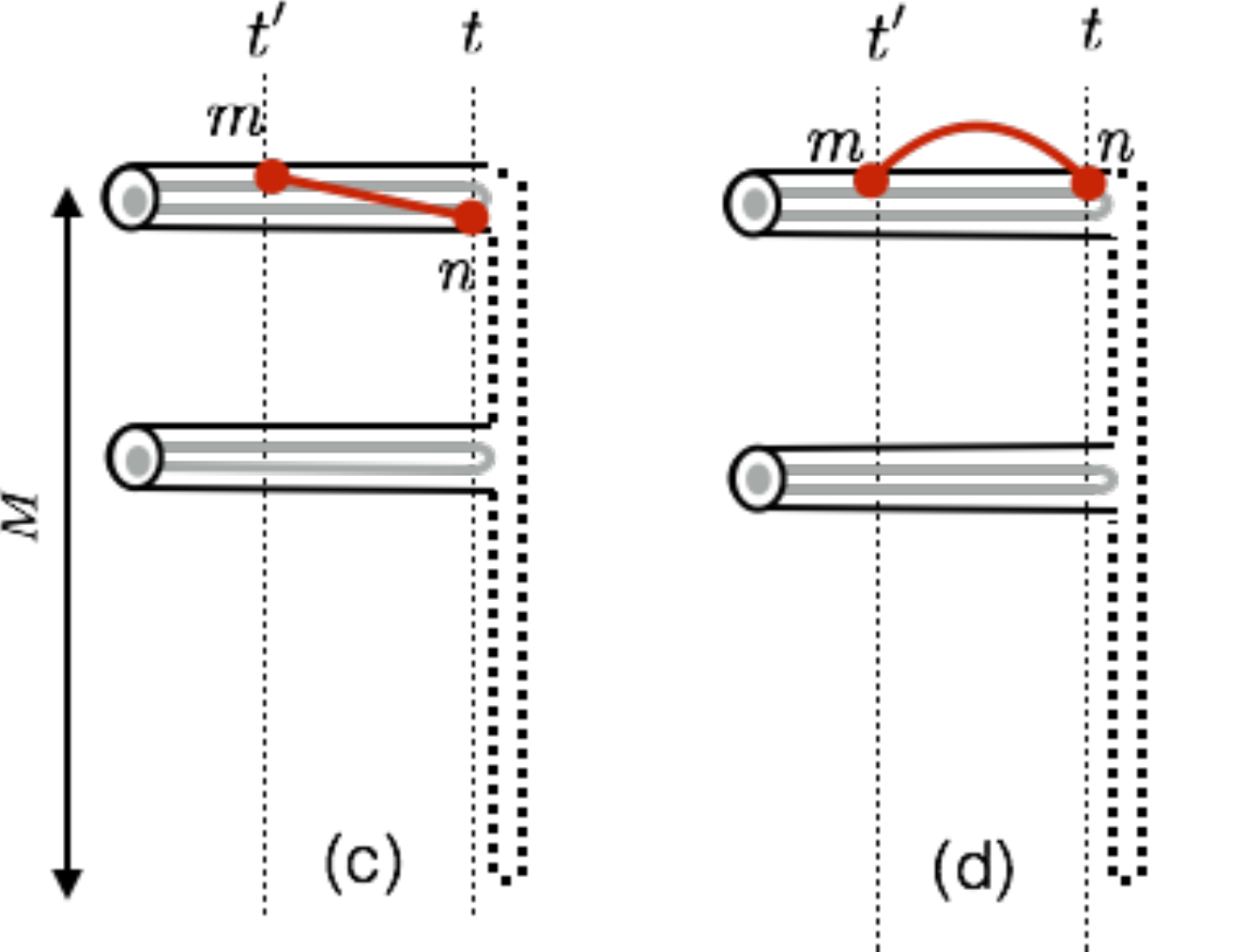}}\caption{(Color online) Multicontour diagrams for the R-flow of exponent $M$ with interaction legs at nodes $m$ and $n$  both in a world. Internal contours (in gray) denote system $Y$ and the outer contours system $X$ (in thermal equilibrium). Interactions take places at vertices $m$ and $n$ at different times one of which is fixed at $t$. At time $t$ the contour of $X$ closes in dotted lines. }
\label{fig. 1w}
\end{center}
\end{figure}


\emph{Single-world R-flow}: The Renyi entropy flows in one-world diagrams are shown in Fig. (\ref{fig. 1w}a-d) : 
\begin{eqnarray} \label{eq. Rflow1w} \nonumber
&& \nonumber \left.\bar{\mathcal{F}}_M\right|_{\textup{1w}}= -{\frac{M}{\tau}\int_{0}^{\tau}dt} \int_{-\infty}^{t}dt' \sum_{mn} \textup{Tr}_{_X} \\ && \nonumber \bigg\{ \frac{  \hat  X_{n}\left(t'\right)\rho_{_X}  \hat  X_{m}\left(t\right)\rho_{_X}^{M-1} -  \hat   X_{m}\left(t\right)  \hat  X_{n}\left(t'\right)\rho_{_X}^{M} }{\textup{Tr}_{_X} \left\{ \rho_{_X}^{M}\right\} }\bigg.   \\&&  \nonumber \qquad \qquad \qquad \qquad \qquad  \ \times \left\langle  \hat Y_{m}\left(t\right)  \hat Y_{n}\left(t'\right) \rho_{_Y}\right\rangle \\ \nonumber
&&+\frac{   \hat  X_{n}\left(t\right)\rho_{_X}  \hat X_{m}\left(t'\right)\rho_{_X}^{M-1} -  \rho_{_X}^{M}  \hat X_{m}\left(t'\right)  \hat X_{n}\left(t\right) }{\textup{Tr}_{_X} \left\{ \rho_{_X}^{M}\right\} } \\ &&  \bigg. \qquad \qquad  \qquad \qquad \qquad   \times\left\langle  \hat  Y_{m}\left(t'\right)  \hat Y_{n}\left(t\right) \rho_{_Y} \right\rangle \bigg\} .
\end{eqnarray}   
 
 The generalised correlators in system $X$ is defined as 
\begin{equation} \label{eq. SNM}
S_{mn}^{N,M}(t-t') \equiv \frac{\textup{Tr}_{_X} \left\{ \hat   X_m(t')\rho_{_X}^N  \hat  X_n(t) \rho_{_X}^{M-N}\right\}}{\textup{Tr}_{_X} \left\{ \rho_{_X}^M\right\} }.
\end{equation}

We generalized the Kubo-Martin-Schwinger (KMS) relation \cite{KMS}   to $M$-worlds  in \cite{AN14}. The Fourier transforming  of generalized correlator thermal in a thermal equilibrium with temperature-independent dynamical susceptibility $\tilde{\chi}_{mn}(\omega)$  can be determined from the relation  (see  Appendix \ref{app3}): 
\begin{equation}\label{eq. KMS}
S_{mn}^{N,M}(\omega)=  \exp({\beta N \omega}) \bar{n}(M \omega/T) \tilde{\chi}_{mn}(\omega) 
\end{equation}
 
 Note that since the time-difference $t-t'$ in the correlator is large we can shift the lower bound of time integral over $t'$ from $0\to -\infty$. 
 
  Note that in a system with temperature-dependent $\tilde{\chi}_{mn}(\omega)$ requires rescaling its temperature to $T/M$. This correlator can be easily shown to be related to the generalized correlators of single-world interactions $S^{0,M}$in the following form:    $S^{N,M}\left(\omega\right)=\exp({\beta N\omega}) S^{0,M}\left(\omega\right)$,
where $S_{mn}^{0,M}\left(\omega\right)=S_{mn}^{(\beta^*)}\left(\omega\right)$  which is the standard spectral density in an environment of rescaled temperature from $T\to T^*=1/k_B\beta^{*}=1/ M\beta$. 

Using these definitions Eq. (\ref{eq. Rflow1w}) is simplified, 
\begin{eqnarray}\label{eq. Rflow1wfinal} \nonumber
&& \left. \bar{\mathcal{F}}_M\right|_{\textup{1w}}  = -{\frac{M}{\tau} \int_{0}^{\tau}dt } \int_{-\infty}^{t}dt'  \\ \nonumber &&   \qquad  \qquad  \sum_{mn}  \int \frac{d\omega}{2\pi}\left(e^{\beta\left(M-1\right)\omega}-1\right)   S_{mn}^{\left(\beta^{*}\right)}\left(\omega\right)  \\ &&\ \ \ \  \times \nonumber    \left\langle  \hat  Y_{m}\left(t'\right)  \hat Y_{n}\left(t\right)e^{-i\omega (t-t')}+  \hat  Y_{m}\left(t\right)  \hat  Y_{n}\left(t'\right)e^{i\omega (t-t')}\right\rangle  \\ 
\end{eqnarray}


\emph{Multi-world R-flow}:  Similarly one can calculate the dynamics associated to multiple-world interactions. In this case the energy  is exchanged between different worlds.   Summing over all possible diagrams and using the generalized KMS relation  to simplify the result the flow of Renyi entropy from different-world interactions can be found: 

\begin{eqnarray}\nonumber \label{eq. Rflowmwfinall}
 && \left. \bar{\mathcal{F}}_M\right|_{\textup{mw}}  = -{\frac{M}{\tau}\int_{0}^{\tau}dt }\int_{-\infty }^{t}dt'\int\frac{d\omega}{2\pi}  \sum_{mn}  \left(e^{\beta\left(M-1\right)\omega}-1\right) \\  && \nonumber   \quad \ \ \times   S_{mn}^{0,M}(\omega)  e^{i\omega (t-t')} \left\langle  \hat Y_{m}\left(t\right)\rho_{Y}\right\rangle \left\langle  \hat  Y_{n}\left(t'\right)\rho_{Y}\right\rangle  \left(e^{\beta\omega}-1\right)\\ &&
\end{eqnarray}
Details of this calculation can be found in Appendix \ref{app 2}.   
 
 The total flow of Renyi entropy can be obtained by summing over Eq. (\ref{eq. Rflow1wfinal}) and (\ref{eq. Rflowmwfinall}). By factorizing terms evolving with the same frequency (i.e. $e^{\pm i\omega (t-t') })$ the final result is 
\begin{eqnarray} \label{eq: final} \nonumber
\bar{\mathcal{F}}_M  &=& - M \sum_{m,n} \int\frac{d\omega}{2\pi}\left(e^{\beta\left(M-1\right)\omega}-1\right)  S_{mn}^{0,M}\left(\omega\right)  \\  && \qquad   \quad  \times {\left( \mathcal{Y}_{mn} (\omega) - \textup{Y}_{mn} (\omega)\right) }
\end{eqnarray}
with $\mathcal{Y}_{mn}$ and $\textup{Y}_{mn}$ defined in Eqs. (\ref{eq. curlyY}) and (\ref{eq. textY}), respectively.

\emph{Correspondence}:  Comparing eq. (\ref{eq: final}) with (\ref{eq. fi}) and (\ref{eq. fc}) one can  conclude the exact correspondence mentioned in Eq. (\ref{eq. corres}).


\emph{The case of Shannon entropy flow}: Let us discuss here how the correspondence look like for the Shannon entropy flow.  The Shannon entropy $S=\textup{Tr} \rho \ln \rho $ can be genuinely defined from the Renyi entropy in the following form: {$S= - \lim_{M \to 1} \partial S_M / \partial M$.} Using the correspondence (\ref{eq. corres}) the flow of Shannon entropy is $\bar{\mathcal{F}}_S=  (i \beta ) \lim_{\xi \to 0} \partial \left( \bar{f}_i - \bar{f}_c \right) / \partial \xi $.   The flow of Shannon entropy exactly corresponds to  

\begin{equation}
\bar{\mathcal{F}}_S^{(\beta)}=  \frac{ Q_{i}^{(\beta^*)}- Q_{c}^{(\beta^*)} 
}{T}  \end{equation} 
with $Q_{i/c}^{(\beta^*)}$ the incoherent and coherent dissipated energy in a system of temperature $T^*=1/k_B\beta^*$ and $\beta^*\equiv M\beta$. \cite{AN14}

 \section{Example 1: The simplest quantum heat engine} 
 
 A quantum heat engine (QHE) is a system of several discrete quantum states connected to several environments at different temperatures. The motivation for research in QHE comes from studying models of photocells and photosynthesis \cite{Scully}. It has been demonstrated that quantum effects can dramatically change the thermodynamics of QHEs \cite{drastic} and their fluctuations \cite{Rahav} manifesting the role of quantum coherence. 
 
The simplest QHE of our interest is made of a probe environment weakly coupled to a two level system (TLS) whose  states are $|0\rangle$ and $|1\rangle$. The TLS itself is also coupled to other heat baths at different temperatures as well as a coherent driving force with frequency $\Omega$ matching the two level energy difference. {The specifics and the simplicity of the situation is that all energy exchanges take place by quanta $\hbar \Omega$.} The interaction between the two level system and the probe  is governed by the interaction Hamiltonian: $ \hat   H_{int}= \hat  X_{01}\left(t\right)|0\rangle\langle1|e^{-i\Omega t}+ \hat  X_{10}(t)|1\rangle\langle0|e^{i\Omega t}$. 

\begin{figure}[htbp] 
\begin{center}
\includegraphics[scale=0.28]{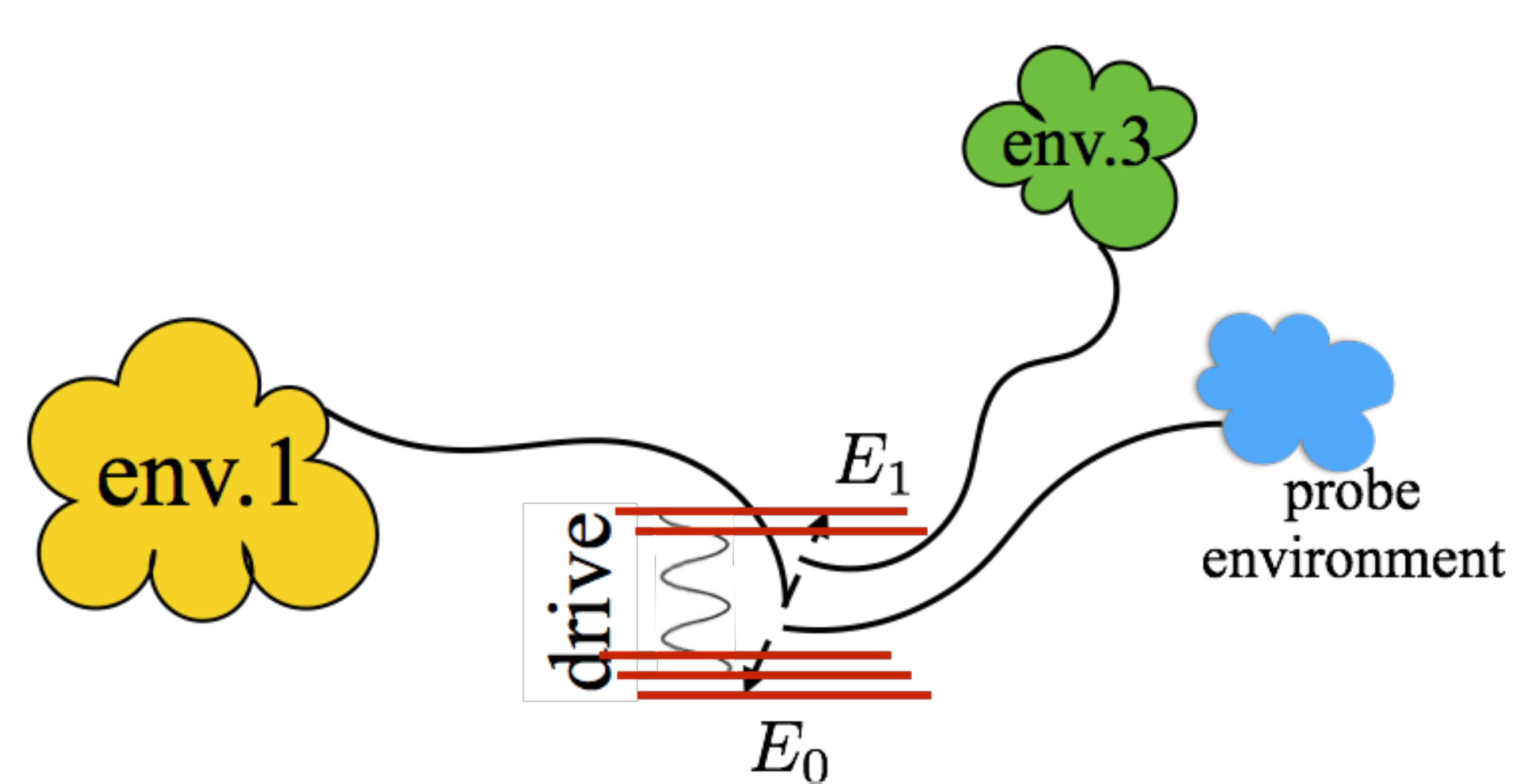}
\caption{(Color online) Schematics of a QHE. A quantum system with two sets of states separated by energy $E_1-E_0$ is driven by external field at matching frequency. The system interacts with a number of environments that induce transitions between the states. We study the R-flows to a weakly coupled probe environment. }
\label{fig. ex2}
\end{center}
\end{figure}

In Ref. \cite{AN14} we explicitly derived the Renyi entropy flow for the probe environment of this system  using perturbative expansion of the probe dynamics. Here we make an attempt to determine the R-flow using the full counting statistics method and the correspondence of Eq. (\ref{eq. corres}). 
  
In the TLS, transition from an upper level to a lower one takes place by the operator $ \hat Y_{10}(t)=|1\rangle\langle0|e^{i\Omega t}$
and the opposite one by $ \hat Y_{01}(t)=|0\rangle\langle1|e^{-i\Omega t}$.
Moreover: $\langle  \hat  Y_{10}(t)  \hat Y_{01}(t')\rho_{s}\rangle=\rho_{11}e^{i\Omega(t-t')}$,
and $\langle  \hat Y_{01}(t) \hat Y_{10}(t')\rho_{s}\rangle=\rho_{00}e^{-i\Omega(t-t')}$.
Also $\langle  \hat Y_{10}\rho_{s}\rangle(t)=\rho_{01}e^{i\Omega t}$ and
$\langle  \hat Y_{01}\rho_{s}\rangle(t)=\rho_{10}e^{-i\Omega t}$.  Let us denote the excited and ground state probabilities with $p_1=\rho_{11}$ and $p_0=\rho_{00}$. 

Using the Kubo-Martin-Schwinger (KMS) relation \cite{KMS} in Eq. (\ref{eq. KMS}) we can introduce the excitation transition rate $\Gamma_{\uparrow}= \bar{n}(\Omega/T ) \tilde{\chi} _{01,10}$, and the emission rate $\Gamma_{\downarrow} =e^{\beta \Omega} \Gamma_{\uparrow}$ with  the Bose  function $\bar{n}(\Omega/T )=1/(e^{\beta \Omega}-1)$.    

Using Eq.(\ref{eq. fi}) the full counting statistics of heat dissipation in the incoherent energy transfer is: 
\begin{equation}  \label{eq. ex1i}
 \bar{f}_i^{(\beta^{*})}\left(\xi^*\right)   = 
  \left( e^{-i\xi^*\Omega}-1 \right)  \frac{\bar{n}(M\Omega/T)}{\bar{n}(\Omega/T)}  \left[\Gamma_\downarrow p_1 -\Gamma_\uparrow p_{0} \right] 
\end{equation}

Similarly,  using Eq.(\ref{eq. fc}), the full counting statistics of energy transfer through quantum coherence flow becomes
\begin{equation} \label{eq. ex1c}
\bar{f}_c^{(\beta^*)}(\xi^*)  = 
  \left( e^{-i\xi^*\Omega}-1 \right)  \frac{\bar{n}(M\Omega/T)}{\bar{n}(\Omega/T)}  (\Gamma_\downarrow - \Gamma_\uparrow ) \rho_{01}\rho_{10}  
\end{equation}
where we used  $S_{mn,pq}(-\omega)=e^{\beta \omega} S_{pq,mn}(\omega)$. 

Notice that Eqs. (\ref{eq. ex1i}) and (\ref{eq. ex1c}) are the FCS associated to  Poisson probabilities.  The reason is that the probe environment is weakly coupled to the quantum heat engine. Since we consider the dynamics to be Markovian, the time lag between two successive emissions in  equilibrium environments at fixed temperatures is long. The events of transmissions of energy are uncorrelated. Moreover due to the weak coupling the energy transfers take place at low transmission probability. Such a process can be described by Poisson probability $p_k=e^{-\bar{n}}\bar{n}^k/k!$ for exchanging $k$ quanta of energy $\hbar \Omega$, where $\bar{n}$ is the average number of quanta transmitted during time $[0,\tau]$.   In the case the coupling of interaction between the two systems is not weak enough, or the emissions take place in short time intervals such that they become  correlated, the Poissonian probability for emissions and absorptions are no longer valid.    

From the correspondence Eq. (\ref{eq. corres}) the R-flow can be obtained from subtracting the two FCSs at $\xi^*$ and $\beta^*$: 
\begin{eqnarray} \label{eq. ex1}\nonumber
\bar{\mathcal{F}}_M^\beta &=&  \frac{M \bar{n}(M\Omega/T)}{\bar{n}((M-1)\Omega/T)\ \bar{n}(\Omega/T)} \times \\&&    \left( p_{1} \Gamma_{\downarrow}-p_{0} \Gamma_{\uparrow}+ \left(\Gamma_{\downarrow}-\Gamma_{\uparrow} \right) \rho_{01}\rho_{10}\right)
\end{eqnarray}

In the second line of Eq. (\ref{eq. ex1}) the first two terms are the dissipation of heat and the third term is the energy transfer due to quantum coherence flow.
In conclusion, what we calculated above matches the R-flow result we obtained earlier in Ref. \cite{AN14}.

\section{ Example 2:  A driven harmonic oscillator coupled to heat baths } \label{secex2}
 
{Let us consider a single harmonic oscillator of frequency $\omega_0$ with Hamiltonian $ \hat  H=\omega_0 (\hat{a}^\dagger \hat{a} +1/2) $ is coupled to a number of environments at different temperatures with different coupling strength. We concentrate on a probe environment which is weakly coupled to the oscillator. In addition the oscillator is driven by external force  at frequency $\Omega$. }

\begin{figure}[htbp] 
\begin{center}
\includegraphics[scale=0.28]{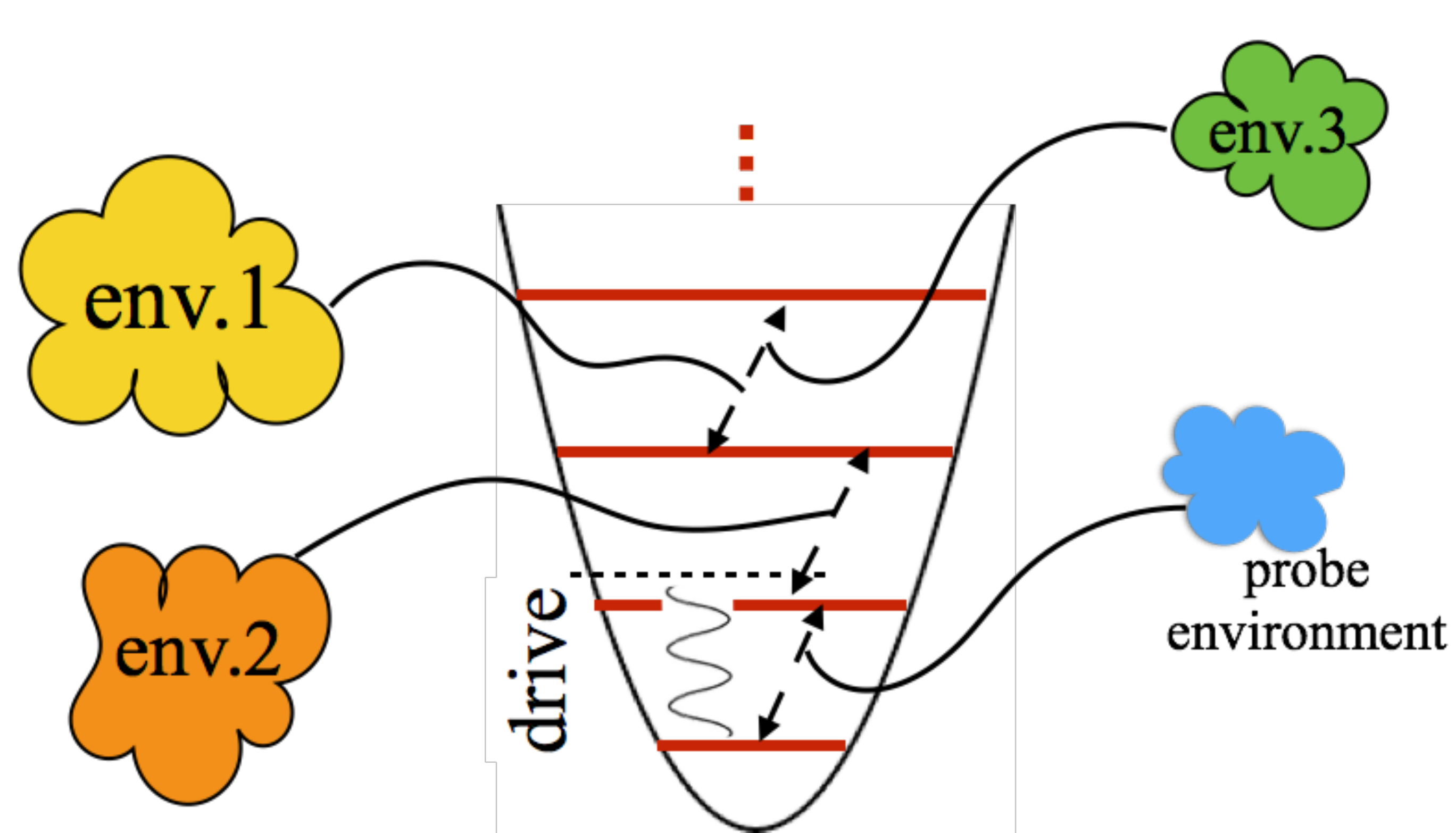}
\caption{Schematics of a harmonic oscillator of frequency $\omega_0$ interacting with environment at different temperatures. The environments   induce transitions between the states in the harmonic oscillator.  The oscillator is driven by external force at frequency $\Omega$.  We study the R-flows to a weakly coupled probe environment. }
\label{fig. ex2}
\end{center}
\end{figure}

We calculate the Renyi entropy flow to the probe environment.  {The coupling Hamiltonian between the harmonic oscillator and the probe reservoir is $\hat{H}(t)= \hat{X}(t) \hat{a}^\dagger(t) +h.c. $ with $\hat{X}$ being the probe reservoir operator. } The Fourier transform of the correlator is: $S_{mn}^{(\beta)}(\omega)=\int \exp(-i\omega t)S_{mn}^{(\beta)}(t) d\omega/2\pi$. Due to conservation of energy the energy exchange occurs either with quantum $\hbar \Omega$ or with quantum $\hbar \omega_0$. 

We note that the time dependence of the average of two operators can be written as $\langle \hat a^\dagger(t) \hat a(t') \rangle = \langle \langle \hat a^\dagger \hat a\rangle \rangle  e^{i\omega_0(t-t')} + \langle \hat a (t)\rangle \langle \hat a^\dagger (t')\rangle  $, where the $\langle a (t) \rangle $ is due to the driving force and therefore oscillates at frequency $\Omega$: $\langle a(t) \rangle = \langle a \rangle_{_+} \exp({i\Omega t}) + \langle a \rangle_{_-} \exp({-i \Omega t})$. This corresponds to the fact that the oscillator can oscillates both at its own frequency and at the frequency of external force.

Obtaining the FCS of energy transfers is straightforward from the diagrams of Fig. (\ref{fig. fcs}). The incoherent and coherent flows are: 
\begin{eqnarray*}\nonumber
- f_i^{(\beta)}(\xi)&=& S^{(\beta )}(\omega_0)\langle \langle a a^\dagger \rangle \rangle (e^{-i\omega_0 \xi}-1)  \\&&     + S^{(\beta )}(-\omega_0) \langle \langle a^\dagger a \rangle \rangle (e^{i\omega_0 \xi}-1)  
\\
& &  + S^{(\beta )}(\Omega) \langle a \rangle_{_-} \langle a^\dagger \rangle_{_+} (e^{-i\Omega \xi}-1)  \\&&    + S^{(\beta )}(-\Omega)\langle a \rangle_{_+} \langle a^\dagger \rangle_{_-} (e^{i\Omega \xi}-1) 
\\
- f_c^{(\beta)}(\xi)&=& S^{(\beta )}(\Omega) \langle a \rangle_{_-} \langle a^\dagger \rangle_{_+} (e^{-i\Omega \xi}-1)  \\&&    + S^{(\beta )}(-\Omega)\langle a \rangle_{_+} \langle a^\dagger \rangle_{_-} (e^{i\Omega \xi}-1) 
\end{eqnarray*}

Substituting these FCSs in the correspondence of Eq (\ref{eq. corres}) using the values of $\xi^*$ and $\beta^*$, the flow of Renyi entropy after using the relation using  $S^{\beta}(-\omega)=\exp({\beta \omega }) S^{\beta}(\omega)$ is determined to:
\begin{eqnarray} \nonumber
\label{eq. HOi}
 \bar{\mathcal{F}}^{(\beta)}_M &=&  M (e^{\beta (M-1)\omega_0}-1)\  S^{(M\beta )}(\omega_0)  \times \\&& \quad   \left\{ \langle \langle a^\dagger a \rangle \rangle   e^{\beta \omega_0} -   \langle \langle  a a^\dagger \rangle \rangle   \right\}
  \end{eqnarray}

Given $T'$ to be the effective temperature of the harmonic oscillator $\langle \langle a a^\dagger \rangle \rangle =\bar{n}(\omega_0/T')+1$ and $\langle \langle a^\dagger a \rangle \rangle =\bar{n}(\omega_0/T')$.    The KMS relation of Eq. (\ref{eq. KMS}) helps to describe the correlator in the thermal bath  in terms of its dynamical susceptibility, i.e.  $S^{(M\beta)}(\omega)=  \bar{n}(M \omega/T) \tilde{\chi}^{(M\beta)}(\omega) $. These help to simplify  Eq. (\ref{eq. HOi}) into:  
\begin{equation} 
\label{eq. HOfinal}
 \bar{\mathcal{F}}^{(\beta)}_M =  \frac{M\bar{n}\left( {M\omega_0}/{T}\right)  \tilde{\chi}}{\bar{n}({(M-1)\omega_0}/{T})\ \bar{n}\left({\omega_0}/{T}\right)}        \left\{ \bar{n}\left({\omega_0}/{T'}\right)   - \bar{n}\left({\omega_0}/{T}\right)       \right\}
  \end{equation}

The entropy flow is robust in the sense that it only depends on the probe and harmonic oscillator temperatures and completely insensitive to external driving force. The entropy flow changes sign  at temperature $T=T'$. \\

 \section{ Discussion}

In this paper we prove  an exact correspondence between the flow of Renyi ( as well as Shannon) entropy and the full counting statistics of energy transfers.  This correspondence is valid for the flow to the system in  thermal equilibrium that is weakly coupled to an arbitrary system out of equilibrium subject to arbitrary time-depending forces.   

In the case of time-dependent external forces we need to introduce an auxiliary full counting statistics  of energy transfers. This is FCS for the case when the quantum forces acting on the system in thermal equilibrium $\hat Y$'s are replaced by their averages. The usual FCS can be in principle measured directly. The same applies to the auxiliary FCS although the measurement protocol is more involved.  Let us describe this protocol. 

Let us notice that the forces correspond to operators $\hat{Y}_n$ and therefore can be in principle measured directly as an expectation value of this observable.  The output of this measurement is a function $\langle \hat{Y}_n(t) \rangle $ which is  periodic with period $\tau$. From this point one can proceed in two ways. First way in to build an artificial system that interacts with system $X$ classically and program it to exert classical forces on system $X$ with values that are given by the results of the first measurement.  One then collects the statistics of energy transfers to obtain the auxiliary FCS. The second way is more practical. One notices that  the response of $\hat{X}_n$ on the forces  is linear one in the limit of weak coupling, so instead of measuring the statistics of energy transfer one can measure the matrix of response functions  $\tilde{\chi}_{mn}(\omega)$. Then the  auxiliary FCS can be evaluated with the aid of Eq. (\ref{eq. fccomp}).

This correspondence allows us to quantify Renyi and Shannon entropy flows. These quantities are not accessible in direct measurement as they are non-linear functions of density matrix.  Direct measurements of density matrix for a probe environment requires characterization of reduced density matrix of an infinite system, which is a rather non-trivial procedure and needs the complete and precise reinitialization of the initial density matrix.   However, measuring the entropy flow from the R/FCS correspondence requires that some generating functions are extracted  from determining statistical cumulants of transferred energy in experimental data. This can be done equally well for imaginary and real values of the characteristic parameter. The measurement procedures may be complex, yet doable and physical.

The correspondence can have many other advantages; for instance: a complete understanding of entropy flows may help to identify the sources of fidelity loss in quantum communications and methods to prevent or control them. 

Our derivation was restricted to the second order perturbative dynamics. There are indications the theorem formulated is not valid in higher orders of perturbation theory. It is interesting to find a similar correspondence that is valid in all order of interaction coupling.

\begin{acknowledgments}
The research leading to these results has received funding from the European Union Seventh Framework Programme (FP7/2007-2013) under grant agreement n° 308850 (INFERNOS).
\end{acknowledgments}

\appendix

\section{A relation} \label{app 1}

 One can easily prove that in general for any multi-argument function $f(\omega, \cdots )$ the following relation holds: 
\begin{eqnarray} \nonumber \label{eq. relation}
&& \int d\omega  \left(e^{\beta (M-1) \omega}-1\right) \ e^{\beta \omega} S_{mn}^{0,M} (\omega) \ f(\omega, \cdots) = \\&&  \nonumber 
- \int d\omega  \left(e^{\beta (M-1) \omega}-1\right) S_{nm}^{0,M} (\omega) \ f(- \omega, \cdots)
\\
\end{eqnarray}

This can be easily proven by changing variable $\omega \to -\omega $ and using the relation between spectral density function of negative and positive frequencies and simplifying using easy algebra.

\section{Generalized KMS} \label{app3}
The generalized correlator of two operators $A$ and $B$ is defined (see eq. (4)):
\begin{equation*}
S^{N,M}_{AB}\left(\omega\right)  =  \int d \tau e^{i\omega \tau} {\rm Tr} \{  \hat A(0) \rho_b^N  \hat  B(\tau) \rho_b^{M-N} \} / {\rm Tr}{\rho_b^M}
\end{equation*}
 
 This correlator in the energy eigenbasis can be rewritten in matrix form

\begin{eqnarray}\nonumber
S_{nm,mn}^{N,M}(\omega) &=&   \int d \tau e^{i\omega \tau}  A_{nm} \frac{e^{-\beta N E_m}}{Z(\beta)^N} \times \\ \nonumber && \ \ \  B_{mn} e^{i (E_m-E_n)\tau} \frac{e^{-\beta E_n (M-N)}}{Z(\beta)^{M-N}}  \frac{Z(\beta)^M}{Z(\beta M)} \\  \nonumber 
&=& 2\pi  \delta\left( E_m-E_n+\omega \right) \frac{A_{nm}  B_{mn} e^{-\beta E_n M} }{Z(\beta M)}e^{\beta N \omega}\\ 
\label{eq. app kms f}
\end{eqnarray}
$Z(\beta)$ is the partition function defined as $Z(\beta)=\sum_i e^{-\beta E_i}$. 

The standard correlator is  $S_{AB}\left(\omega\right)  =  \int d \tau e^{i\omega \tau} {\rm Tr} \{ A(0)  B(\tau) \rho_b \}/ {\rm Tr}{\rho_b}$ becomes equal to 
\begin{equation*}
S_{AB}\left(\omega\right)=  2\pi  \delta\left( E_m-E_n+\omega \right) A_{nm}  B_{mn} e^{-\beta E_n } /Z(\beta )
\end{equation*}
where KMS relation links this to dynamical susceptibility: $S_{AB}(\omega)=\tilde{\chi}_{AB}(\omega)\bar{n}(\omega/T)$. Substituting this in (\ref{eq. app kms f}) a generalized KMS relation is obtained:

\begin{eqnarray}
S^{N,M}_{AB}\left(\omega\right) & = & \bar{n}\left(M\omega/T\right) e^{\beta\omega N} \tilde{\chi}_{AB}\left(\omega\right)\label{eq. app kms1}
\end{eqnarray}

\section{Multiple-world dynamics} \label{app 2}

\begin{figure}[H]
\begin{center}
\centerline{\includegraphics[width=0.5\linewidth]{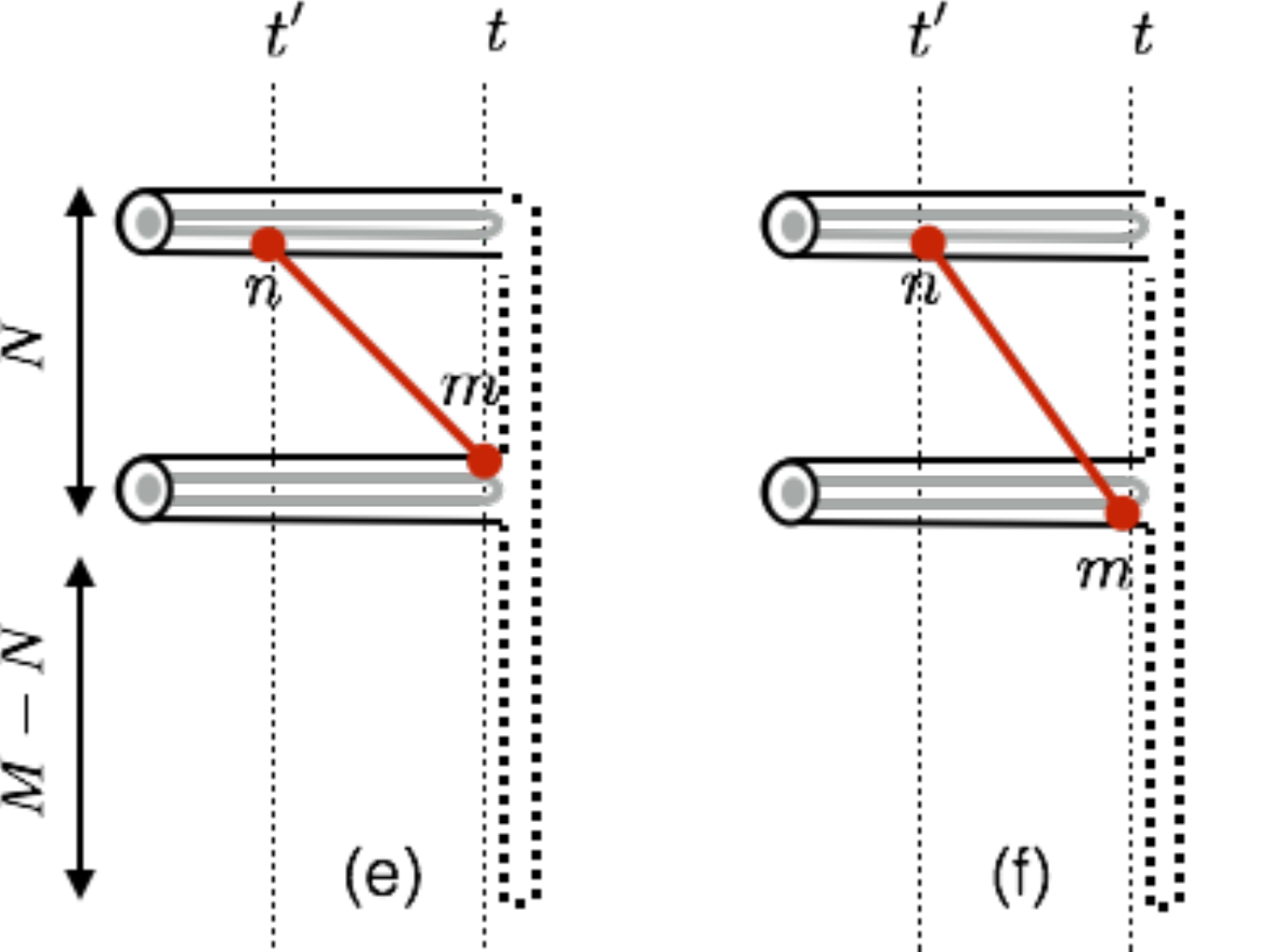}\includegraphics[width=0.5\linewidth]{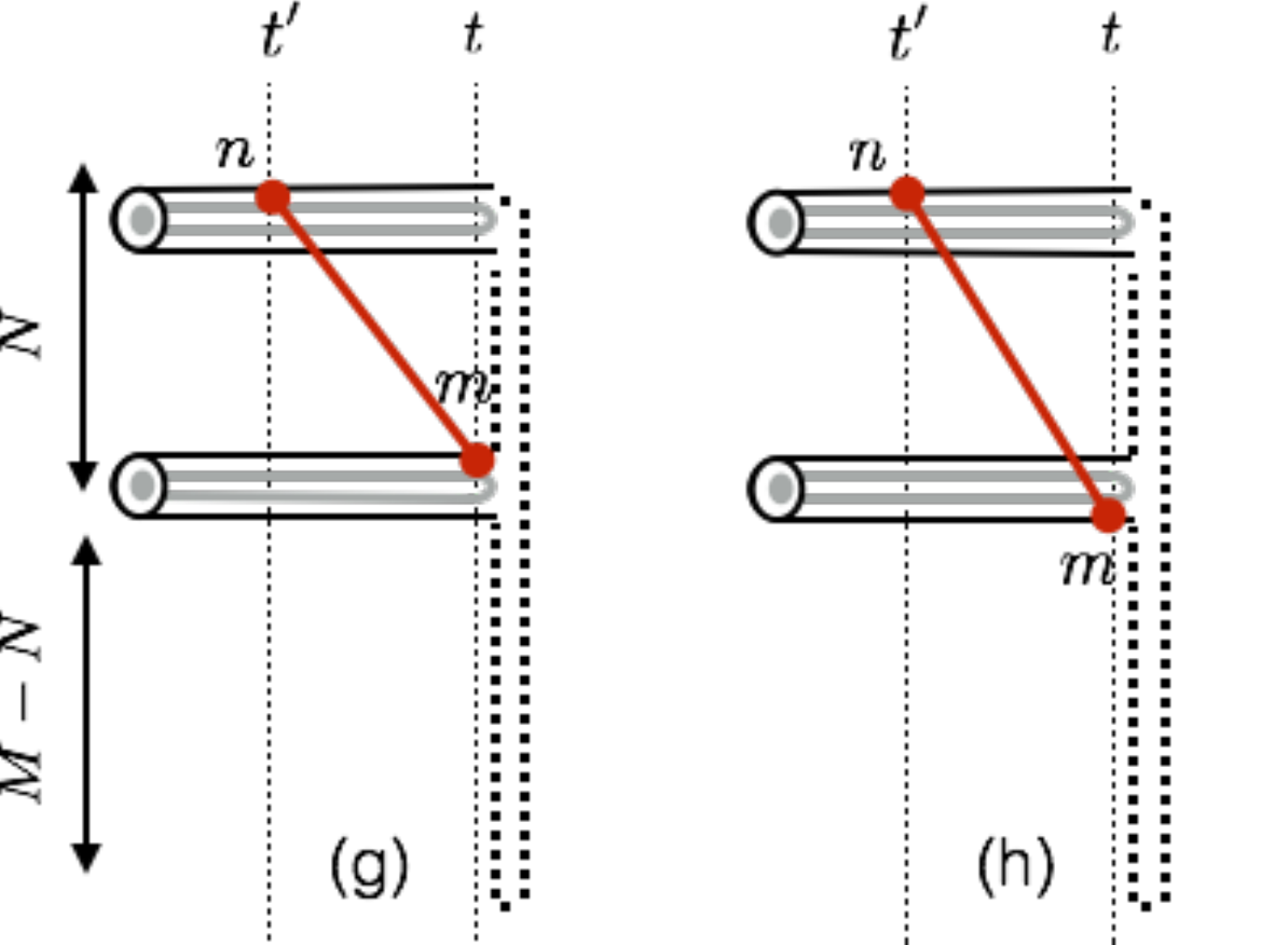}}\centerline{\includegraphics[width=0.5\linewidth]{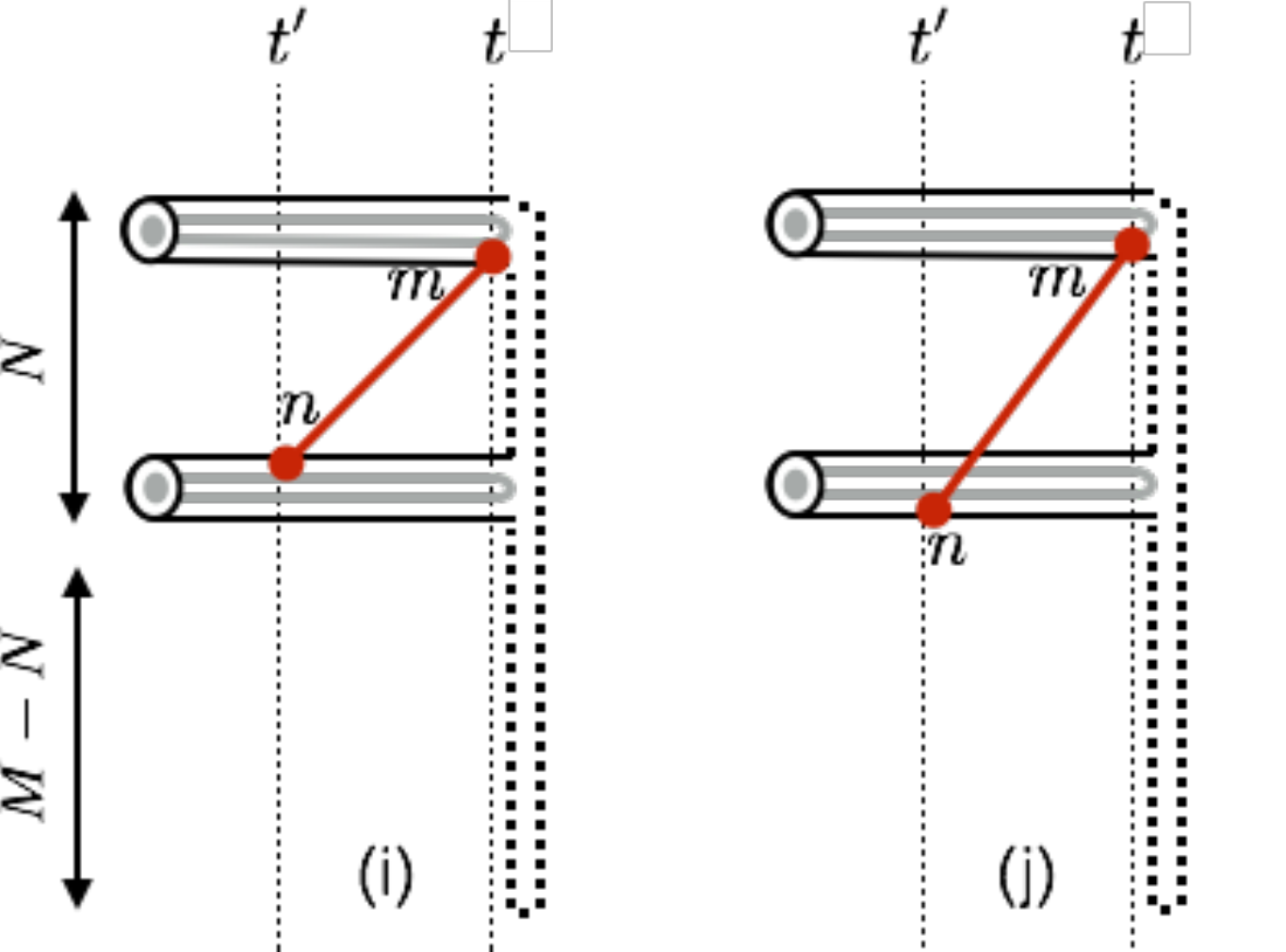}\includegraphics[width=0.5\linewidth]{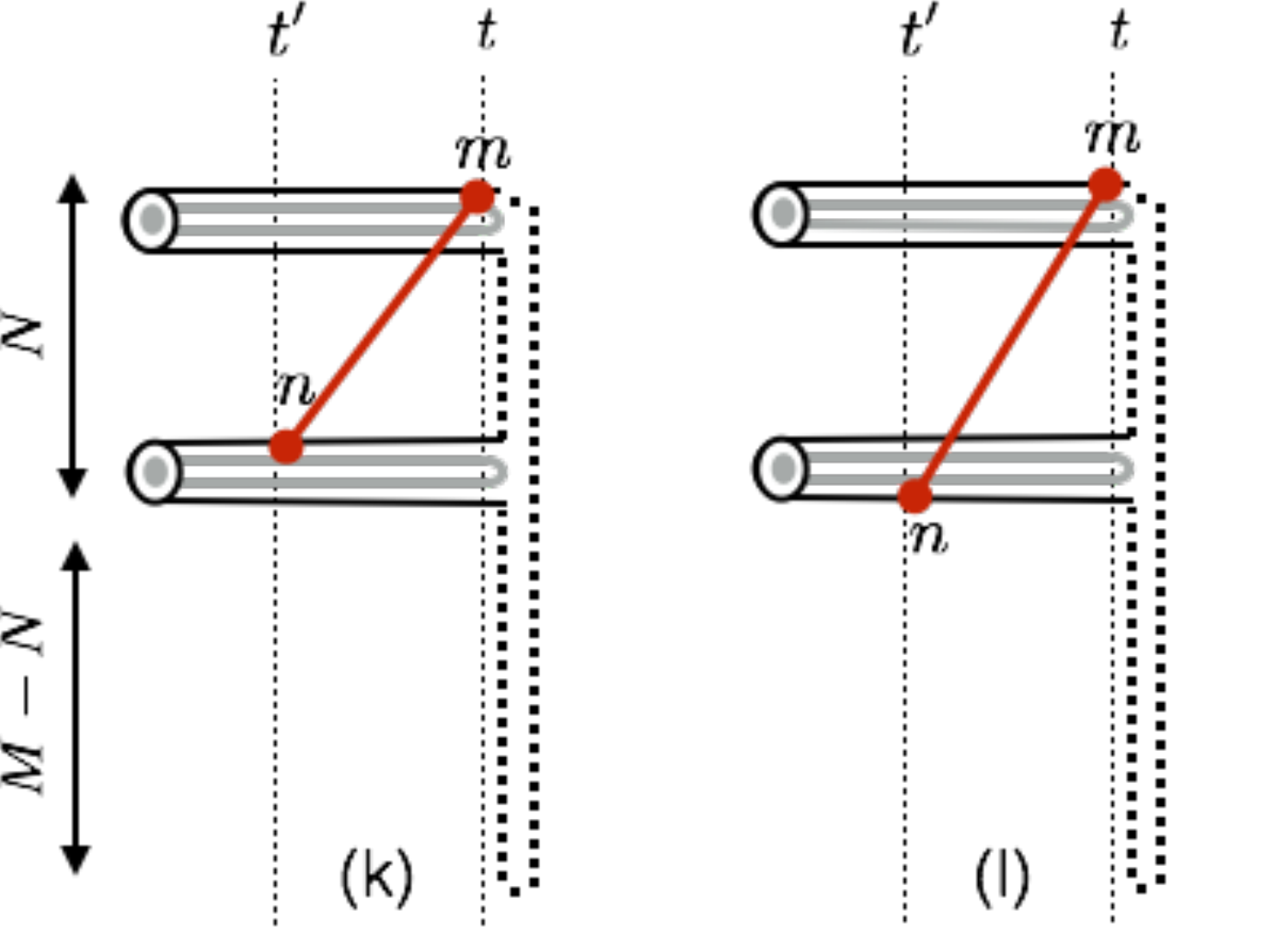}}
\caption{(Color online) Typical multicontour diagrams for the R-flow with interaction legs at nodes $m$ and $n$ in two different worlds. }
\label{fig. mw}
\end{center}
\end{figure}

Typical diagrams corresponding to the multi-world terms are listed in Fig. (\ref{fig. mw})

Similar to the analysis for one world, we can write detailed diagrammatic values for the evolution of R-flow in the case two interactions occurs in two different worlds. After Fourier transformation the flow from the diagrams (e-l) of Fig. (\ref{fig. mw}) becomes:
\begin{eqnarray*} \label{eq. Rflowmw1}
&& \left. \bar{\mathcal{F}}_M\right|_{\textup{(e-l)}}   =  -\frac{1}{\tau}\int_{0}^{\tau}dt \int_{- \infty}^{t}dt'\int\frac{d\omega}{2\pi} \sum_{mn} \\&& \qquad \qquad \times   \left\langle  \hat Y_{m}\left(t\right)\rho_{Y}\right\rangle \left\langle  \hat Y_{n}\left(t'\right)\rho_{Y}\right\rangle \times \\&& 
\left\{ e^{i\omega (t-t')}\left[S_{mn}^{N-2,M}\left(\omega\right)-2S_{mn}^{N-1,M}\left(\omega\right)+S_{mn}^{N,M}\left(\omega\right)\right] \right. \\&& \left.+e^{-i\omega (t-t')}\left[S_{nm}^{N-2,M}\left(\omega\right)-2S_{nm}^{N-1,M}\left(\omega\right)+S_{nm}^{N,M}\left(\omega\right)\right]\right\} 
\end{eqnarray*}

This must be summed over all possibilities. When the first interaction is at the topmost world the second one can run between $n=2$ and $M$. However, when we put the first interaction at the second topmost world the second interaction can have maximally $M-1$ world distance with it, therefore $n=2$ to $M-1$. Note that we already consider the both positive and negative energy exchanges in the summation of diagrams (e-l). Extending this discussion one can find the following total summation for all multi-world diagrams: 

\begin{eqnarray*} \label{eq. Rflowmw}
&& \left. \bar{\mathcal{F}}_M\right|_{\textup{mw}}   =  -\frac{1}{\tau}\int_{0}^{\tau}dt \int_{- \infty}^{t}dt'\int\frac{d\omega}{2\pi} \sum_{mn} \\&& \qquad \qquad \times  \sum_{M'=2}^M \sum_{N=2}^{M'} \left\langle  \hat Y_{m}\left(t\right)\rho_{Y}\right\rangle \left\langle  \hat  Y_{n}\left(t'\right)\rho_{Y}\right\rangle \times \\&& 
\left\{ e^{i\omega (t-t')}\left(S_{mn}^{N-2,M}\left(\omega\right)-2S_{mn}^{N-1,M}\left(\omega\right)+S_{mn}^{N,M}\left(\omega\right)\right) \right. \\&& \left.+e^{-i\omega (t-t')}\left(S_{nm}^{N-2,M}\left(\omega\right)-2S_{nm}^{N-1,M}\left(\omega\right)+S_{nm}^{N,M}\left(\omega\right)\right)\right\} 
\end{eqnarray*}

Changing $\omega \to -\omega$ in terms with indices $S_{nm}$ and using the relation $S_{AB}^{N,M}(-\omega)= S_{BA}^{M-N,M}(\omega)$ that can be easily concluded from the Fourier transforming eq. (\ref{eq. SNM}) and simplifying the summation using the KMS relation, all multi-world diagrams sum into
\begin{eqnarray}\nonumber \label{eq. Rflowmwfinal}
 && \left. \bar{\mathcal{F}}_M\right|_{\textup{mw}}  = -\frac{M}{\tau}\int_{0}^{\tau}dt \int_{-\infty }^{t}dt'\int\frac{d\omega}{2\pi}  \sum_{mn}  \left(e^{\beta\left(M-1\right)\omega}-1\right) \\  && \nonumber   \quad \ \ \times   S_{mn}^{0,M}(\omega)  e^{i\omega (t-t')} \left\langle  \hat Y_{m}\left(t\right)\rho_{Y}\right\rangle \left\langle \hat Y_{n}\left(t'\right)\rho_{Y}\right\rangle  \left(e^{\beta\omega}-1\right)\\ &&
\end{eqnarray}

Simplifying the integration using a relation that comes in Eq. (\ref{eq. relation})  we can further simplify this relation into 

\begin{equation} \label{eq. mw}
 \left. \bar{\mathcal{F}}_M\right|_{\textup{mw}}  = M \int\frac{d\omega}{2\pi}  \sum_{mn}  \left(e^{\beta\left(M-1\right)\omega}-1\right) 
    S_{mn}^{0,M}(\omega)  \textup{Y}_{mn}(\tau, \omega)
\end{equation}

\end{document}